\documentclass[10pt,aps,prl,twocolumn,superscriptaddress]{revtex4-2}
\pdfoutput=1 
\usepackage{microtype}
\usepackage{times}
\usepackage{amsfonts}
\usepackage{mathrsfs}
\usepackage{amsmath}
\usepackage{amssymb}
\usepackage{fancyhdr}
\usepackage{graphicx}
\usepackage{xspace}
\usepackage{rotating}
\usepackage[normalem]{ulem}
\usepackage{braket}
\usepackage{verbatim}
\usepackage{xcolor}
\usepackage[utf8]{inputenc}
\usepackage[medium]{titlesec}
\usepackage{bm}
\usepackage[normalem]{ulem}
\usepackage{extarrows}
\usepackage{slashed}
\usepackage{isodateo}
\usepackage{graphicx}
\usepackage{xcolor}
\usepackage[bookmarksnumbered=true,bookmarksopen=true]{hyperref}
\hypersetup{colorlinks,%
            linkcolor=[rgb]{0,0.0,1.0}, %
            citecolor=[rgb]{0,0.0,1.0}, %
            urlcolor=[rgb]{0,0.1,0.9}}
\usepackage[hmargin=.7in,vmargin=1.1in]{geometry}
\usepackage{indentfirst}
\usepackage{cancel}
\usepackage{subcaption}

\makeatletter
\newcommand\footnoteref[1]{\protected@xdef\@thefnmark{\ref{#1}}\@footnotemark}
\makeatother

\usepackage{amsfonts}
\usepackage{mathtools}
\usepackage{pifont}
\usepackage{mathrsfs}
\usepackage{amsmath}
\usepackage{amssymb}
\usepackage{framed}

\newcommand{\benum}{\begin{enumerate}}
\newcommand{\eenum}{\end{enumerate}}
\newcommand{\bi}{\begin{itemize}}
\newcommand{\ei}{\end{itemize}}

\newcommand{\beq}{\begin{equation}}
\newcommand{\eeq}{\end{equation}}

\newcommand{\bea}{\begin{eqnarray}}
\newcommand{\eea}{\end{eqnarray}}

\newcommand{\Rmnum}[1]{\expandafter\@slowromancap\romannumeral #1@}

\def\bga{\begin{aligned}}
\def\eda{\end{aligned}}
\def\bgp{\begin{pmatrix}}
\def\edp{\end{pmatrix}}
\def\bgs{\begin{subequations}}
\def\eds{\end{subequations}}

\def\to{\rightarrow}

\newcommand{\dx}{\mathrm{d}}

\renewcommand{\rm}{\mathrm}

\definecolor{cardinal}{rgb}{0.77, 0.12, 0.23}
\newcommand{\prlsection}[1]{\textit{#1~---}}

\begin{document} 

\title{
Seeing in the Dark: Gravitational Waves from  Resonant Production of Dark Photon Dark Matter
}
\author{Yanou Cui}
\email[]{yanou.cui@ucr.edu}
\affiliation{Department of Physics and Astronomy, University of California, Riverside, CA 92521, USA}

\author{Pankaj Saha}
\email{pankaj.saha@apctp.org}
\affiliation{
Asia Pacific Center for Theoretical Physics (APCTP), Pohang 37673, Korea
}
\affiliation{Institute of Particle and Nuclear Studies~(IPNS),\\High Energy Accelerator
Research Organization (KEK), Oho 1-1, Tsukuba 305-0801, Japan}

\author{Evangelos I. Sfakianakis}
\email{evangelos.sfakianakis@austin.utexas.edu}
\affiliation{
Texas Center for Cosmology and Astroparticle Physics,
Weinberg Institute for Theoretical Physics, Department of Physics,
The University of Texas at Austin, Austin, TX 78712, USA}
\affiliation{Department of Physics, Harvard University, Cambridge, MA, 02131, USA}

\date{\today}

\begin{abstract}
Ultralight dark photons (DPs) are well-motivated dark matter candidates. A promising yet underexplored mechanism for their relic abundance is parametric resonance in the Abelian-Higgs model that generates the DP mass. We advance previous analytical studies through dedicated 3D lattice simulations that capture important nonlinear effects, including backreaction and relic partitioning between the dark Higgs and DPs. We further predict the associated stochastic gravitational-wave (GW) signal, which can be within reach of a broad range of detectors, opening a new avenue to probe ultralight DP dark matter.

\end{abstract}
\maketitle
\prlsection{Introduction.} 
Dark photons (DPs) are gauge bosons of a hidden Abelian symmetry that naturally arise in a broad class of extensions to the  Standard Model (SM)~\cite{Holdom:1985ag,Redondo:2008ec,Arias:2012az,Graham:2015rva,Alexander:2016aln}. Depending on their mass, DPs can mediate interactions with a dark sector or constitute all or part of dark matter (DM)~\cite{Redondo:2008ec,Arias:2012az,Graham:2015rva,Alexander:2016aln}. In the ultralight regime, $m_{A'}\ll$~eV, DP DM is particularly compelling: it is cosmologically stable, and behaves as a coherent, wave-like field targeted by a broad experimental program~\cite{Caputo:2021eaa}.

Since a thermal population would be far too warm, ultralight DP DM must be produced nonthermally~\cite{ParticleDataGroup:2026mpi}, and a variety of such mechanisms have been proposed~\cite{Arias:2012az,Graham:2015rva,Ema:2019yrd,Kolb:2020fwh,Nakai:2020cfw,Agrawal:2018vin,Co:2018lka,Bastero-Gil:2018uel, Adshead:2023qiw, Dror:2018pdh,Long:2019lwl, Kitajima:2022lre}.
A particularly economic mechanism relies on the DP mass being generated by a dark Higgs, as naturally occurs in an Abelian-Higgs model. 
If the dark Higgs is initially displaced from its late-time minimum at the end of inflation, its subsequent oscillations give the vector a periodically varying mass and can produce DPs by parametric resonance. This mechanism was originally studied in detail in Ref.~\cite{Dror:2018pdh}, where DP relic abundance is inferred using linear Floquet theory. 
However, linear analysis alone cannot capture important aspects such as backreaction, rescattering, and the energy distribution of residuals, potentially missing key phenomenological consequences.

In this Letter, we advance our understanding of this underexplored mechanism through dedicated three-dimensional lattice simulations capturing the nonlinear Abelian-Higgs dynamics following resonant Higgs-vector production. We demonstrate that the resonance does not merely populate DPs: it fragments the dark Higgs condensate, producing an inhomogeneous post-resonance state and determining the energy partition between DPs and dark Higgs excitations. The resonant production also sources a stochastic gravitational-wave (GW) background, whose peak frequency traces the characteristic energy scale of the resonance and thereby probes the mass scale of the DP sector. The dark matter relic abundance, late-time dark-sector composition, and GW signal thus provide complementary probes of the same underlying nonperturbative and nonlinear dynamics.

The rest of the Letter is organized as follows. 
In Sec.~II we introduce the minimal Abelian-Higgs benchmark and the large-field quartic regime.
In Sec.~III we summarize the linear resonance structure and demonstrate the effects of nonlinear fragmentation based on lattice simulations. 
In Sec.~IV we calculate the post-resonance energy partition and dark matter relic abundance. 
In Sec.~V we compute the stochastic GW signal and show how it correlates with the energy scales in the
DP DM sector. Finally, we conclude in Sec.~VI.

\bigskip

\par
\prlsection{Benchmark Model.} 
We consider the minimal gauged Abelian Higgs dark model where the gauge field (DP) acquires mass by the Higgs mechanism, and can be efficiently produced through parametric resonance.
The model consists of a complex scalar $\Phi$
charged under $U(1)_X$ and a DP $A_\mu$. The dark-sector Lagrangian density is
\begin{equation}
    \mathcal{L}_{X} = (D_{\mu}\Phi)^{\ast}(D^{\mu}\Phi) - V(\Phi^{\ast},\Phi) -\frac{1}{4}F_{\mu\nu}F^{\mu\nu},
    \label{eq:lag}
\end{equation}
where $D_{\mu} = \nabla_{\mu} + i ({g_A})A_{\mu}$ is the covariant derivative and $F_{\mu\nu} = \nabla_{\mu}A_{\nu} - \nabla_{\nu}A_{\mu}$ is the DP field strength tensor. 
A gauge-invariant kinetic mixing term $\epsilon F_{\mu\nu}F_Y^{\mu\nu}/2$ may be present, where $F_Y^{\mu\nu}$ is associated with the SM gauge group $U(1)_Y$. 
The constraints and detection prospects associated with kinetic mixing depend on the DP mass and model details, and have been extensively studied in the literature~\cite{Cline:2024qzv,Caputo:2026pdw}. 
In our scenario, $\epsilon$ is not required to establish the DP relic abundance and cannot be generated radiatively within the minimal model setup. 
Thus, $\epsilon$ may consistently vanish, allowing the model to readily evade existing constraints from kinetic mixing. 
The presence of a small nonzero $\epsilon$, while not required in our setup, would offer complementary laboratory and astrophysical probes of dark photons beyond the mass-dependent searches considered here. A detailed analysis of this possibility is beyond the scope of this work.

We consider the minimal symmetry-breaking quartic potential, inspired by the SM,
\begin{equation}
V(\Phi^\ast,\Phi)=-\mu^2|\Phi|^2+\lambda_\Phi|\Phi|^4 .
\label{eq:dark-higgs-potential}
\end{equation}
By expressing the scalar in terms of real fields as
$
\Phi=\frac{1}{\sqrt{2}}\left(\varphi_R+i\varphi_I\right)$,
the late-time vacuum can be chosen to locate at $v\equiv \langle \varphi_R\rangle={\mu}/{\sqrt{\lambda_\Phi}}$, 
leading to the late-time masses
$
m_A={g_A v}$ 
 and
$
m_h=\sqrt{2}\,\mu$. Here $h$ denotes the radial excitation mode of $\Phi$. 
 In the absence of a bare mass term for $A_\mu$ in $\mathcal{L}_{X}$, the action is invariant under the gauged $U(1)_X$ and the DP mass is entirely generated by the spontaneous breaking of the $U(1)_X$.

We consider the Abelian–Higgs model minimally coupled to gravity on a spatially flat Friedmann–Lema\`itre–Robertson–Walker (FLRW) background. 
The covariant equations of motion following from the Lagrangian in Eq.~\eqref{eq:lag} are
\begin{subequations}
\label{eq:EOM}
\begin{align}
D_\mu D^\mu\Phi+\frac{\partial V}{\partial \Phi^\ast} &=0 \, ,
\label{eq:EqPhi}\\
\nabla_\nu F^{\nu\mu}=J^\mu &=2\,g_A\,\mathrm{Im}\!\left(\Phi^\ast D^\mu\Phi\right)\,, 
\label{eq:EqA}
\end{align}
\end{subequations}
where Eq.~\eqref{eq:EqA} corresponds to the dark-sector Maxwell's equations.


A central ingredient of the mechanism is the large initial displacement of the dark Higgs from its late-time vacuum at the end of inflation, $|\Phi_{\rm {in}}|\gg v$.
In this large-field regime, the quadratic term in Eq.~\eqref{eq:dark-higgs-potential} is negligible, and the dynamics is governed by the quartic interaction. 
Thus, we will refer to this regime as the ``quartic regime.'' The large initial displacement can generally arise for a light spectator field during inflation~\cite{Starobinsky:1994bd}. 
For a quartic spectator potential, the de Sitter equilibrium estimate gives
$\sqrt{\langle |\Phi_{\rm {in}}|^2\rangle}\sim {\cal O}(1)\, {H_{\rm I}}/{\lambda_\Phi^{1/4}}$.
Thus a Planckian displacement, $|\Phi_{\rm {in}}|\sim M_{\rm {Pl}}$, can be obtained for $H_{\rm I}\sim \lambda_\Phi^{1/4}M_{\rm {Pl}}$, where $H_{\rm I}$ stands for the Hubble scale at the end of inflation. 
This stochastic spectator realization requires more than the observable $\sim 60$ e-folds to reach equilibrium.
We remain agnostic about the origin of $\Phi_{\rm {in}}$, treating it phenomenologically as an initial condition for the subsequent radiation-era evolution.

After inflation, during radiation domination, the field remains overdamped while $H\gg m_{\mathrm{eff}}$. 
In the quartic regime, $m_{\mathrm{eff}}^2\sim \lambda_\Phi |\Phi_{\rm{in}}|^2$, therefore the condensate begins to roll towards the true vacuum and oscillate when $H_{\rm {osc}}\sim \sqrt{\lambda_\Phi}\,|\Phi_{\rm {in}}|$.
For $|\Phi_{\rm {in}}|\sim M_{\rm {Pl}}$, this gives $H_{\rm {osc}}\sim \lambda_\Phi^{1/2}M_{\rm {Pl}}<H_I$.

\bigskip

\prlsection{Resonance and Dark Photon Production.}
During the subsequent oscillatory stage, the quartic-dominated Abelian-Higgs theory in a radiation-dominated universe is conformally invariant. 
Upon using conformal time $d\tau=dt/a$ and rescaling the scalar field as $\chi=a\Phi$, the scale factor drops out of the large-field quartic dynamics because $a''=0$, where $'\equiv d/d\tau$. 
Consequently, the envelope of the physical condensate amplitude redshifts as $|\Phi|\propto a^{-1}$.  
The radial part of the rescaled field $\chi$ oscillates with an approximately constant amplitude. 
Consequently, $a^2 m_A(\tau) = g_A^2 |\chi(\tau)|^2$ has a constant envelope but periodic time dependence.

For the linear resonance analysis, it is convenient to work in Coulomb
gauge, $\partial_i A_i=0$. 
The spatial gauge field is then
purely transverse and contains the two transverse polarizations
$A_\pm$, while the longitudinal physical degree of freedom is encoded
in the phase fluctuation of the complex scalar.
Choosing the homogeneous condensate to lie along the real
direction, we write
$
\Phi=\frac{1}{\sqrt{2}}
\left(\varphi_R+\delta\varphi_R+i\delta\varphi_I\right )$,
so that $\delta\varphi_I$ describes the Goldstone/longitudinal sector. The
temporal component $A_0$ is non-dynamical and is fixed by Gauss's law.
Observable quantities, such as energy densities and gravitational-wave
spectra, are gauge independent~\cite{Lozanov:2016pac}. The Coulomb-gauge variables provide a
convenient description of the resonance.

The linearized equation for the transverse polarizations in Coulomb gauge is
\beq
\partial_\tau^2 A^\pm_{k} + \left (
k^2 + a^2m_A^2(\tau) 
\right ) A^\pm_{k}=0 \, ,~ m_A^2(\tau) = g_A^2 \phi^2(\tau)\, .
\label{eq:Apm}
\eeq
The periodic time dependence of $a m_A(\tau)$ drives parametric resonance
without Hubble friction in conformal variables.

The resultant resonance dynamics associated with the Higgsed vector (DP), including the distinct transverse and longitudinal mode equations, was studied in~\cite{Dror:2018pdh} based on the linear approximation. Here we first follow the linear Floquet picture to identify the unstable bands and then capture the subsequent nonlinear fragmentation with lattice simulations.

Integrating out $A_0$ and canonically normalizing the longitudinal mode via the field redefinition $\delta\phi_{I,k} = \sqrt{1+{a^2m_A^2}/{k^2}}\, \delta\widetilde{\phi}_{I,k}$ yields a standard oscillator equation:
\begin{equation}
\label{eq:Aleom}
\partial_\tau^2\delta\widetilde{\phi}_{I,k}
+
\omega_L^2
\delta\widetilde{\phi}_{I,k}
=0 \, ,
\end{equation}
with the effective frequency
\begin{equation}
\omega_L^2
=
k^2+a^2m_A^2
+
\frac{
3\left[\partial_\tau(a m_A)/k\right]^2
}{
\left(1+\frac{a^2m_A^2}{k^2}\right)^2
}
-
\frac{
\partial_\tau^2(a\phi_R)/(a\phi_R)
}{
1+\frac{a^2m_A^2}{k^2}
}\, .
\label{eq:omegaL}
\end{equation}

Following Eqs.~\eqref{eq:Apm},~\eqref{eq:Aleom} and~\eqref{eq:omegaL},
the transverse and longitudinal gauge modes can   be analyzed using Floquet theory~\cite{Dror:2018pdh, Lozanov:2016pac}.

The  resonance of the gauge field  (transverse and longitudinal DP modes) is controlled by the dimensionless resonance parameter $q \equiv {g_A^2}/{\lambda_\Phi}$.
The quartic interaction also induces radial self-resonance of the dark Higgs. For $q \gtrsim 1$, vector production is more efficient than radial self-resonance of the dark Higgs (i.e., the gauge-dominated regime), while $q \gg 1$ corresponds to broad resonance. The characteristic physical momenta can be written 
as $k_{\mathrm{phys}} \sim \kappa_{\rm{res}}(q)
\sqrt{\lambda_\Phi}\,|\Phi|$,
or, in conformal variables,
$k \sim \kappa_{\rm{res}}(q)
\sqrt{\lambda_\Phi}\,a|\Phi|$,
where $\kappa_{\rm{res}}(q)$ is a dimensionless factor determined by the resonance structure.

To quantify the instability, we define the dimensionless Floquet exponent
$\mu_k$ through
$A_k(z) \propto e^{\mu_k z}$,
where $z$ is conformal time measured in units of the background oscillation frequency. Linear analysis gives $\mu_k \simeq 0.1$--$0.2$ for the most
unstable gauge modes, whereas radial self-resonance is weaker, with $\mu_k \lesssim 0.03$ in the gauge-dominated regime~\cite{Dror:2018pdh, Lozanov:2016pac, longpaper}.

However, the linear description breaks down once the produced gauge and scalar fluctuations backreact on the homogeneous condensate. At this point, the system enters a genuinely nonlinear regime: the coherent dark Higgs condensate fragments, while its energy is redistributed among gauge and scalar modes, and large spatial gradients generate anisotropic stress. This stage determines both the late-time partition of the dark-sector relic abundance and the stochastic GW signal. One must therefore evolve the full Abelian-Higgs system on a lattice beyond the linear Floquet regime. 

We have developed a custom code to solve Eqs.~\eqref{eq:EOM} in the temporal gauge in $3+1$ dimensions. Unlike most previous studies, we evolve the Hubble rate self-consistently with the dark-sector energy density and pressure through the Friedmann equation, rather than assuming a prescribed background evolution~\cite{longpaper}. This ensures a consistent treatment of the subsequent GW redshifting and proper normalization relative to the critical energy density.

Our simulations show that the resonance excites both transverse and longitudinal gauge degrees of freedom. Backreaction also induces radial dark Higgs fluctuations, with the gauge and scalar fields typically carrying comparable amounts of energy, $\rho_{\rm {Higgs}}/\rho_{\rm {gauge}}={\cal O}(1)$. 
In contrast, linear Floquet analysis, as used in earlier literature, predicts an exponential difference between scalar and gauge modes.
This contrast demonstrates the importance of backreaction and the necessity of lattice simulations for a proper characterization of the final state of the dark sector. 
Furthermore, the residual homogeneous condensate is strongly depleted; at the end of
our simulations we find
${|\langle\Phi\rangle|}/
{\sqrt{\langle|\Phi|^2\rangle}}
\sim 10^{-3}$.
Thus the post-resonance state is not a coherent homogeneous condensate, rather a highly inhomogeneous configuration containing gauge-field fluctuations, scalar fluctuations, and large spatial gradients, with the scalar sector becoming phase-disordered. The total anisotropic stress of this inhomogeneous Abelian-Higgs system thus sources the stochastic GW background in a very generic way.

An important feature of the quartic regime is that its radiation-like scaling persists beyond the linear analysis. Neglecting the quadratic scale $\mu$, the Abelian-Higgs action contains no dimensionful parameters and remains conformally invariant during radiation domination, even after backreaction and fragmentation. Consequently, the total dark-sector energy density redshifts as $\rho_X\propto a^{-4}$. Once the resonance shuts down and the comoving energy partition largely stabilizes, the gauge fields, scalar fluctuations, and any residual coherent component follow the same scaling, up to residual interactions.

\bigskip

\prlsection{Fragmentation, Post-Resonance Dynamics, and Dark Matter Relic Abundance.}
The cosmological history of the dark sector proceeds through four dynamical phases: 
(i) Initially, the dark Higgs is frozen at a large, super-Hubble displacement.
(ii) As the Universe expands and cools, Hubble friction decreases and the dark Higgs field begins coherent oscillations in the quartic region of its potential, driving explosive parametric resonance of the gauge field. 
(iii) The resonance rapidly fragments the coherent dark Higgs condensate into a highly inhomogeneous, phase-disordered, nonthermal state. During this phase the spatial average of the complex field is strongly suppressed, $\langle\Phi\rangle\ll  \phi_{\rm {rms}}\equiv \sqrt{\langle|\Phi|^2\rangle}$, while the local rms amplitude remains large, $\langle|\Phi|^2\rangle\gg v^2$. The gauge field therefore propagates in an inhomogeneous dark Higgs background with an
effective mass scale set by the local variance, $ m_A^2(\tau)\sim g_A^2\langle|\Phi|^2\rangle$.
We refer to this as a locally Higgsed, nonthermal state. It is distinct from a thermally restored symmetric phase in the high temperature early Universe, which is commonly considered in the literature for a Higgs model. In our scenario, the dark sector is sequestered from the SM bath, and was never in a symmetric phase. This is a \textit{novel yet rather generic possibility} of the cosmic evolution of an Abelian Higgs sector. 
In the quartic regime the entire fragmented dark sector continues to redshift as radiation. 
(iv) Finally, as the Universe  cools further, the local variance redshifts until it
reaches the true vacuum: 
$\langle|\Phi|^2\rangle\simeq v^2={\mu_\Phi^2}/{\lambda_\Phi}$,
at which point the system becomes trapped in the symmetry-broken vacuum,
and the DP and dark Higgs acquire their constant late-time
masses.

 After trapping, the Higgs vacuum fixes the particle spectrum in the model: $m_h=\sqrt{2}\,\mu$ and $m_A=\sqrt{q/2}\,m_h$, where $q\equiv g_A^2/\lambda_\Phi$. The low-momentum modes then behave as nonrelativistic matter. Crucially, because the physical momenta of the Higgs fluctuations redshift as $1/a$ in lockstep with the background field amplitude, their typical momenta at vacuum trapping (denoted as ``vt") are $k_{\mathrm{vt, Higgs}} \sim \sqrt{\lambda_\Phi} v$. Consequently, for a gauge-dominated resonance ($q \gg 1$), when the Higgs settles into its minimum, the massive DP is already non-relativistic ($k_{\mathrm{vt, gauge}} \ll m_A$), whereas the massive dark Higgs at the time of vacuum trapping is at the threshold between relativistic and non-relativistic. As long as vacuum trapping occurs at least a few $e$-folds before matter-radiation equality ($t_{\rm{eq}}$), both species behave as cold dark matter at $t_{\rm{eq}}$. 
 To make this statement quantitative, in the following we estimate $T_{\rm vt}$ by tracking the local rms field amplitude,
$\phi_{\rm{rms}}$. 

At a reference epoch $\tau_\ast$, during the quartic regime, we define the dark-sector energy fraction as
\begin{equation}
\label{eq:aPhistar}
\alpha_{\Phi,\ast}\equiv \frac{\rho_{X,\ast}}{\rho_{\mathrm{tot},\ast}}
= \frac{\rho_{X,\ast}}
{\rho_{X,\ast}+\rho_{\mathrm{SM},\ast}} \, .
\end{equation}

During the quartic-dominated era, $\phi$ energy scales like radiation, $\phi_{\rm {rms}}\propto a^{-1}$. Using $\rho_X\simeq \alpha_{\Phi,*}\rho_{\rm{tot}}\sim \lambda_\Phi \phi_{\rm{rms}}^4$, we find $\phi_{\rm {rms}}(T)\sim \lambda_\Phi^{-1/4}\alpha_{\Phi,*}^{1/4}T$. Vacuum trapping of the field $\phi$ occurs when $\phi_{\rm {rms}}\sim v=\mu/\sqrt{\lambda_\Phi}$, yielding
$
T_{\rm{vt}}\sim \lambda_\Phi^{-1/4}\alpha_{\Phi,*}^{-1/4}\mu$. The aforementioned cold dark matter condition requires $T_{\rm{vt}}\gtrsim T_{\mathrm{eq}}\simeq0.8\,{\mathrm{eV}}$, which leads to the bound
$\mu\gtrsim \lambda_\Phi^{1/4}\alpha_{\Phi,*}^{1/4}T_{\mathrm{eq}}$. 

In the gauge-dominated regime of interest ($q\gtrsim1$), the decay $h\to AA$ is kinematically forbidden, therefore the leading interaction between $h$ and $A$ after vacuum trapping is the number-conserving process $AA\leftrightarrow hh$. 
Considering ${m_h}/{m_A}=\sqrt{{2}
/{q}}$, 
we find that for $1/2<q<2$ the direction $hh\to AA$ is kinematically favored, with $AA\to hh$ being favored  for $q>2$.
Summing over the yields stored in DP fluctuations, radial Higgs fluctuations, and the (highly suppressed) residual homogeneous condensate, we obtain the dark-sector relic abundance (for a detailed derivation see Section~\ref{sec:relic} of the Supplemental Material:
\begin{align}
\nonumber
\Omega_{\rm X}h^2
\simeq
6.1\,
\mathcal{D}_N\,\Xi_N\,F_m
&\left(\frac{g_*}{100}\right)^{3/4}\!
\left(\frac{100}{g_{*s}}\right)\!
\left(\frac{10^{-40}}{\lambda_\Phi}\right)^{1/4}\\
&\times
\alpha_{\Phi,*}^{3/4}
\left(\frac{m_A}{10^{-8}\,{\mathrm{eV}}}\right).
\label{eq:OmegaX}
\end{align}

The dimensionless factor $\Xi_N\sim{\cal O}(1)$
sums up all three energy compositions, and is extracted from the end-state of our lattice simulations.
In our simulations, $\Xi_N\sim{\cal O}(1)$ is read directly from the computed energy and number partition among the three components, rather than inferred from linear Floquet growth rates, which break down once backreaction and rescattering become important~\footnote{
One might argue that, since $\Xi_N\sim\mathcal{O}(1)$, linear Floquet estimates should suffice. However, both the $\mathcal{O}(1)$ magnitude of $\Xi_N$ and its precise value are determined by the nonlinear evolution. Floquet theory captures the growth rates in the linear regime, but the energy partition is established at backreaction, while the momentum distribution --- and hence the coldness of the relic --- is determined by subsequent rescattering. Lattice simulations are therefore required not merely to refine an overall normalization, but to determine the phase-space information, encoded in $\kappa_i^{E,p}$ and $r_A^{N,f}$, on which both the relic abundance and coldness constraint depend. These quantities are intrinsically beyond the scope of linear theory.
}. 
The general factor $\mathcal{D}_N$ captures potential post-resonance changes in the total comoving number. For us $\mathcal{D}_N=1$, since $AA\leftrightarrow hh$ conserves the total dark-sector number.

The composition factor $F_m=r_{A}^f+\frac{m_h}{m_A}(1-r_{A}^f)$ is the mass-weighted relic composition in units of $m_A$, where 
$r_{A}^f\equiv Y_A^f/(Y_A^f+Y_h^f)$ is the  dark-photon number fraction after the parametric resonance has shut down and $m_h/m_A=\sqrt{2/q}$ is the scalar-to-vector mass ratio. 
We used the superscript ``$f$'' to denote final (post-vacuum-trapping) values. Efficient number-conserving scattering can redistribute the population between the two species. The mass hierarchy determines the kinematically favored direction. Therefore, we have
\begin{equation}
F_m\simeq
\begin{cases}
1, & r_{A}^f\!\to\!1\ \ (\text{vector-dominated}) \, ,\\
\sqrt{2/q}, & r_{A}^f\!\to\!0\ \ (\text{scalar-dominated}) \, .
\end{cases}
\label{eq:Fq-main}
\end{equation}
The upper case holds when the DM population is initially dominated by DPs and redistribution is inefficient or drives $hh\to AA (q<2)$; the lower when efficient $AA\to hh$ converts the heavier vector to lighter scalars ($q \geq 2$), leading to $F_m \to m_h/m_A = \sqrt{2/q}$.

Eq.~\eqref{eq:OmegaX} reproduces the parametric yield of resonant DP production~\cite{Dror:2018pdh}, $Y_A\propto\lambda_\Phi^{-1/4}\alpha_{\Phi,*}^{3/4}$, with the proportionality constant and the gauge-scalar energy partition now fixed by the nonlinear evolution and encoded in $\Xi_N$, instead of being left as undetermined phenomenological parameters. 
For a detailed the derivation of the yield, see~\cite{longpaper} and Section~\ref{sec:relic} of the Supplemental Material.

\begin{figure*}[t]
    \centering
    \begin{minipage}[t]{0.48\textwidth}
        \centering
        \includegraphics[width=\linewidth]{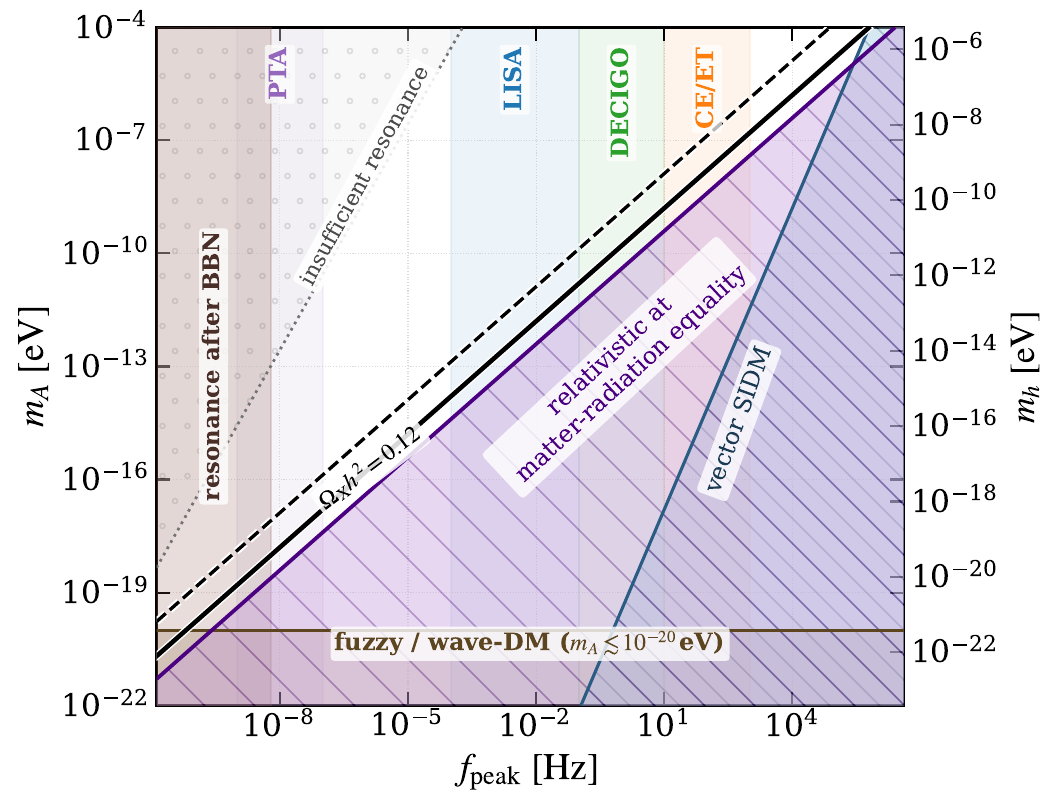} 
    \end{minipage}%
    \hfill
    \begin{minipage}[t]{0.48\textwidth}
        \centering
        \includegraphics[width=\linewidth]{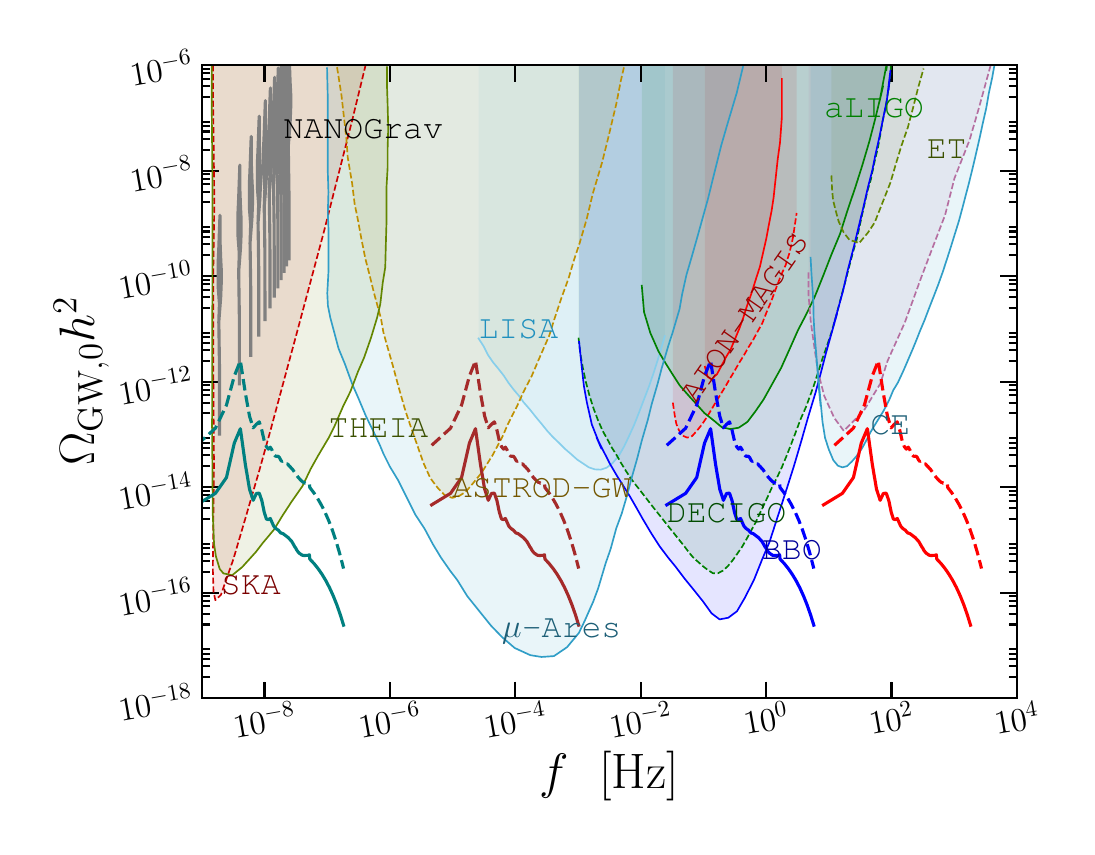} 
    \end{minipage}

    \vspace{0.3cm} 

    \begin{minipage}[t]{0.48\textwidth}
        \raggedright 
        \captionof{figure}{
Relic abundance contours and relevant constraints in the $(f_{\rm{peak}},m_A)$ plane, with our benchmark values overlaid on the sensitivity regions of various GW detectors in the corresponding frequency bands.
The solid and dashed black lines show the relic contour $\Omega_{\rm X}h^2=0.12$ for negligible post-simulation redistribution and maximal conversion respectively, corresponding to the upper and lower cases of Eq.~\eqref{eq:Fq-main}.
Partial redistribution yields a relic abundance between these two limits.
The shaded regions denote the constraints detailed in the main text and Supplemental Material.
        }
        \label{fig:plotRelic}
    \end{minipage}%
    \hfill
    \begin{minipage}[t]{0.48\textwidth}
        \raggedright 
        \caption{
Benchmark examples of present-day stochastic GW spectra from nonlinear fragmentation of the dark Higgs condensate, computed from our lattice simulations. 
Following Eq.~\eqref{eq:mAvsf}, which relates $f_{\rm{peak}}$ to $m_A$,  
the colored GW spectra  correspond, from high to low frequency, to $m_A\sim10^{-8},\,10^{-11},\,10^{-14},
\,10^{-18}\,{\rm eV}$ (red, blue, brown, green). 
Solid and dashed curves correspond to initial dark sector energy fraction $\alpha_{\Phi,*}=1\%$ and $5\%$ respectively.
Shaded regions indicate projected sensitivities of representative GW experiments. }
        \label{fig:plotGWs}
    \end{minipage}
\end{figure*}

The relic density today is shown in Fig.~\ref{fig:plotRelic} as contours on the $(f_{\rm peak}, m_A)$ plane. $f_{\rm{peak}}$ is the GW peak frequency predicted from our benchmark models, which will be elaborated in the next section (see Eq.~\eqref{eq:fpeak_fit}).
The  relic density depends on the final partition among the constituents.  We can identify the two limiting cases: \emph{no conversion}, for which the composition remains fixed at its  value at the time of vacuum trapping, and $F_m$ takes its largest value (not necessarily $1$), and \emph{maximal conversion}, for which $r_A^{f}\to0$ and $F_m\to\sqrt{2/q}=m_h/m_A$. 
The actual result generally lies between these limits.
The colored vertical bands illustrate the sensitivities of representative GW detectors. The y-axis on the right is $m_h=\sqrt{2/q}\,m_A$, where we assume the benchmark example $q=1.4\times 10^3$, which ensures effective resonance. 
The $\Omega_\chi h^2=0.12$ line lies safely in a regime where the dark sector has become cold relic by the time of matter-radiation equality. 
Constraints from the self-interacting vector dark matter (SIDM) only become relevant for GW frequencies above $\sim {\rm {MHz}}$~\cite{Markevitch:2003at,
Randall:2008ppe,Tulin:2017ara}. Furthermore, the constraint on fuzzy DM~~\cite{Hu:2000ke,Hui:2016ltb,Rogers:2020ltq,DES:2020fxi} and the requirement that parametric resonance occurs before BBN~~\cite{Planck:2018vyg,
Fields:2019pfx} only affect the part of parameter space relevant for pulsar-timing arrays~\cite{NANOGrav:2023gor,
EPTA:2023fyk,Reardon:2023gzh,Xu:2023wog}. Finally, black hole superradiance can constrain certain regions of DP masses below $\sim 10^{-11}\,$eV~\cite{Baryakhtar:2017ngi,Cardoso:2018tly}.
However, for a Higgsed DP as in our case, the formation of
vortex strings may halt the growth of the superradiance clouds~\cite{East:2022ppo,East:2022rsi}. These bounds thus apply only for sufficiently large Higgs vacuum energy~\cite{Jones:2024fpg,Aswathi:2025nxa}, a condition not met by our benchmarks.

\bigskip

\prlsection{Gravitational Wave Signature.}
The nonlinear fragmentation of the dark Higgs condensate generates large anisotropies in scalar and gauge fields. 
The resulting total anisotropic stress of the Abelian-Higgs system sources a stochastic GW background with a peaked spectrum. 
The tensor modes evolve according to
\begin{equation}
h_{ij}''+2{\cal H}h_{ij}'-\nabla^2h_{ij}
=
16\pi G a^2\Pi_{ij}^{\rm{TT}},
\label{eq:hij_eom}
\end{equation}
where \({\cal H}=a'/a\), primes denote derivatives with respect to
conformal time \(\tau\), and \(\Pi_{ij}^{\rm{TT}}\) is the
transverse-traceless projection of the full scalar-plus-gauge anisotropic
stress.

Using the GW energy density 
$
\rho_{\mathrm{GW}} = 
\left\langle h'_{ij}h'_{ij}\right\rangle / (32\pi G a^2)$
we compute the dimensionless GW energy spectrum
\begin{equation}
\Omega_{\mathrm{GW}}(k,\tau) = \frac{1}{\rho_{\mathrm{tot}}}
\frac{d\rho_{\mathrm{GW}}}{d\ln k}
=
\frac{k^3 \int d\Omega_k
\left\langle
h'_{ij}(\mathbf{k},\tau)
h'^*_{ij}(\mathbf{k},\tau)
\right\rangle}{32\pi G a^2 (2\pi)^3 \rho_{\rm{tot}}}\, .
\label{eq:diml_GW_spec}
\end{equation}
Here $d\Omega_{\hat k}$ denotes the angular integral over directions of $\mathbf{k}$.

After the anisotropic-stress source becomes negligible, the produced GWs propagate freely, with $\rho_{\rm {GW}}\propto a^{-4}$. 
Hence, the ratio $\rho_{\rm{GW}}/\rho_{\rm{rad}}$ remains constant during radiation domination as long as the relativistic degrees of freedom are fixed. The subsequent change in the visible sector relativistic degrees of freedom transfers entropy to the thermal bath but not to the GWs, leading to the standard entropy-dilution factor.

We denote the end of the lattice evolution by $\tau_{\rm{end}}$, which is chosen late enough so that the GW spectrum has stabilized. We choose $\tau_{\rm{end}}$ after the production epoch, which occurs near $H_{\rm{osc}}$. 
For simplicity, we take $g_*=g_{*s}$ at the production epoch.
Here $g_*$ and $g_{*s}$ denote the effective relativistic degrees of freedom in the SM sector, contributing to the energy and entropy densities, respectively.
Redshifting the spectrum from $\tau_{\rm{end}}$ to today gives the following result:
\begin{equation}
\Omega_{\mathrm{GW},0}(k_0)
\simeq
\Omega_{\mathrm{rad},0}
\left(\frac{g_{*,0}}{g_{*,\mathrm{end}}}\right)^{1/3}
\Omega_{\mathrm{GW}}(k,\tau_{\mathrm{end}}).
\end{equation}
The corresponding present-day frequency is
\begin{equation}
f_0
=
\frac{k}{2\pi a_0}
=
\frac{k_{\rm {phys,*}}}{2\pi}
\frac{T_0}{T_*}
\left(\frac{g_{s,0}}{g_{s,*}}\right)^{1/3} \, ,~
k_{\rm {phys,*}}\equiv \frac{k}{a_*} \, .
\label{eq:f0_GW}
\end{equation}

The characteristic scale of the spectrum is fixed by the time of production. 
The GW emission is the strongest during the nonlinear fragmentation stage, after the onset of the resonance, so the relevant Hubble scale $H_*$ is smaller but close to $H_{\rm {osc}}$. 
We can use $H_*\sim H_{\rm{osc}}$ for estimates and determine the precise peak frequency from our lattice simulations. 
In the quartic regime, where 
$V(\phi) \approx \lambda_\Phi |\Phi|^4$,
$
H_{\rm {osc}}\sim \sqrt{\lambda_\Phi}\,\phi_{\rm {in}}$ and
$k_{\rm{phys,*}}\sim \kappa_{\rm {res}}(q)\sqrt{\lambda_\Phi}\,\phi_{\rm {in}}
$.
Hence
$
f_{\rm{peak},0}
\propto
\kappa_{\rm{res}}(q)\,
\lambda_\Phi^{1/4}
\left({\phi_{\rm in}}/{M_{\rm {Pl}}}\right)^{1/2} $.
For the benchmark value \(q=1.4\times10^3\), our simulation results are well fitted by 
\begin{equation}
f_{\rm {peak,0}}
\simeq
320\,{\rm{Hz}}\,
\left(\frac{\lambda_\Phi}{10^{-40}}\right)^{1/4}
\left(\frac{\phi_{\mathrm{in}}}{M_{\rm {Pl}}}\right)^{1/2} \, .
\label{eq:fpeak_fit}
\end{equation}
We expect the peak value to depend mildly on $q$; a detailed parameter scan will be presented in~\cite{longpaper}.

For the benchmark $q\sim 10^3$, Eq.~\eqref{eq:fpeak_fit} maps the GW band probed by
ground and space-based interferometers onto very small quartic couplings. 
For $\phi_{\rm{in}}\sim M_{\rm {Pl}}$, the ground-based interferometer band corresponds to  $\lambda_\Phi\sim 10^{-50}-10^{-42}$ for $f_{\rm{peak,0}}\sim 1-100\,{\rm{Hz}}$, while PTA frequencies correspond to 
$\lambda_\Phi\sim10^{-86}-10^{-78}$. 
Such $\lambda_\Phi\ll 1$ is technically natural, and exponentially small self-couplings can arise in UV constructions such as deformed-CFT/dilaton-radion models~\cite{Agrawal:2016ubh}.

For fixed $q$ and $\alpha_{\Phi,*}$, the associated dark-sector masses are then obtained by imposing the relic abundance in Eq.~\eqref{eq:OmegaX}. 
Figure~\ref{fig:plotRelic} shows the resulting target range, bounded by the two limiting post-trapping histories: frozen number fractions and efficient $AA\leftrightarrow hh$ redistribution to the lighter species. 
In the LIGO/CE/ET band, this corresponds to ultralight DPs in the sub-$\mu{\rm{eV}}$ range for the benchmark shown.

To give a numerical example for the $f_{\rm {peak},0}$-$m_A$ relation, we may consider  $\alpha_{\Phi,\ast}\sim{\cal O}(1\%)$, $\phi_{\rm {in}}\sim \rm{Mp}$ and $q\gg 1$, while 
assuming the DP sector accounts for all the dark matter relic density. We then obtain a simple relation between the GW peak frequency and the DP mass by combining Eq.~\ref{eq:OmegaX} and \ref{eq:fpeak_fit} 
\beq
f_{\rm {peak},0}\sim \frac{m_A}{10^{-10}\rm {eV}} \rm{Hz} \, .
\label{eq:mAvsf}
\eeq

The peak amplitude is determined by the initial dark-sector energy fraction and by the nonlinear efficiency of GW production. 
For the benchmark spectra shown in Fig.~\ref{fig:plotGWs}, after
redshifting to today, the peak amplitude can be estimated by
\beq
\left.\Omega_{\rm{GW},0}h^2\right|_{\rm {peak}}
\simeq
10^{-9}\alpha_{\Phi,*}^2 .
\eeq

The parameter $\alpha_{\Phi,*}$ is potentially subject to BBN constraints. 
We found that for the light mass range of our interest (in terms of $\mu$ or $m_A$) as required by the detectable GW frequency range, vacuum trapping (at $T_{\rm{vt}}$ as introduced earlier) of the $\phi$ field typically has not yet occurred by the BBN epoch. 
Thus, during the BBN era, the dark sector remains relativistic and contributes to the radiation density. 
We may characterize its effect in the form of its contribution to the effective number of neutrino species 
$\Delta N_{\mathrm{eff}}^{\mathrm{BBN}}
\equiv
\frac{\rho_X(T_{\mathrm{BBN}})}{\rho_{\nu,1}(T_{\mathrm{BBN}})}$ with
$\rho_{\nu,1}=
\frac{7}{4}\frac{\pi^2}{30}T_{\mathrm{BBN}}^4$. $\rho_X$ denotes the actual dark-sector energy
density. 
Evaluating this relation before electron--positron annihilation, where
$g_*^{\mathrm{BBN}}=43/4$, gives
$\Delta N_{\mathrm{eff}}^{\mathrm{BBN}}
=
\frac{43}{7}
{\rho_X}/{\rho_{\gamma}}$.
This relation does not require the dark and visible sectors to have the
same temperature. 
Since the fragmented dark sector is nonthermal, we do not assign it a temperature or an effective number of thermal degrees of freedom. Its contribution to BBN is characterized directly by its radiation-like energy density $\rho_X$. Relating the ratio $\rho_X / \rho_{\rm {SM}}$ to the initial fraction
$\alpha_{\Phi,*}$ gives
\begin{equation}
\Delta N_{\mathrm{eff}}^{\mathrm{BBN}}
\simeq
\frac{43}{7}
\frac{\alpha_{\Phi,*}}{1-\alpha_{\Phi,*}}
\mathcal{G}(T_{\mathrm{BBN}},T_*),
\end{equation}
where
\begin{equation}
\mathcal{G}(T_{\mathrm{BBN}},T_*)
\equiv
\frac{g_*(T_*)}{g_*^{\mathrm{BBN}}}
\left[
\frac{g_{*s}^{\mathrm{BBN}}}{g_{*s}(T_*)}
\right]^{4/3}
\end{equation}
accounts for the entropy transfer in the visible sector between $T_*$ and $T_{\rm {BBN}}$. For $T_*\gtrsim T_{\rm {BBN}}$, one finds  $0.46\lesssim {\cal G}\le 1$. 
Using the largest value $\mathcal{G}=1$ and the bound $\Delta N_{\mathrm{eff}}^{\mathrm{BBN}}\lesssim0.3$, comparable to a $2\sigma$ allowance from BBN analysis~\cite{Fields:2019pfx}, implies $\alpha_{\Phi,*}\lesssim0.05$ (smaller ${\cal G}$ allows for larger $\alpha_{\Phi,*}$).
This sets the upper bound on $\alpha_{\Phi,*}$ as shown in Fig.~\ref{fig:plotGWs}, and therefore the highest peak GW amplitude. Smaller $\alpha_{\Phi,*}$ gives weaker signals that may remain within the reach of future interferometers over a portion of the parameter space.

Fig.~\ref{fig:plotGWs} illustrates viable, benchmark examples of GW signals from the scenario we consider. As can be seen, the signals are well within reach of GW detectors over a wide frequency range.

Combining the GW spectrum with the relic-abundance relation in
Eq.~\eqref{eq:OmegaX} makes the scenario particularly predictive. The GW peak
frequency identifies the scale of the dark Higgs fragmentation, while the peak amplitude
constrains $\alpha_{\Phi,*}$. 
Assuming this dark sector accounts for all the observed dark matter abundance further fixes the required vacuum trapping scale, which corresponds to the characteristic energy scale of the dark sector, such as the DP and dark Higgs masses. 
This, in turn, demonstrates a strong correlation between the GW spectrum and key properties of the dark sector.

\bigskip

\prlsection{Conclusion.} 
The production of dark photon dark matter via parametric resonance in a dark Abelian-Higgs model is an important yet underexplored mechanism. 
We studied the fully nonlinear evolution associated with this mechanism, using dedicated three-dimensional lattice simulations. 
Our work sheds light on essential physics not captured in earlier studies, such as the fragmentation of the dark Higgs condensate, including gauge backreaction, scalar and vector rescattering, depletion of the coherent condensate, and the gravitational-wave signal sourced by the full Abelian-Higgs anisotropic stress. 
The same nonlinear evolution determines the yield and momentum distribution of dark-sector particles at the end of the parametric resonance epoch, including the partitioning between dark photons and dark Higgs excitations. 
Thus, by capturing the nonlinear dynamics, our lattice simulation enables a self-consistent determination of both the resultant dark matter relic abundance and the associated GW signal, quantities that the analytic linear Floquet treatment~\cite{Dror:2018pdh} could constrain at best up to undetermined $\mathcal{O}(1)$ factors. 

Our study uncovered that the late-time relic population of the dark Abelian Higgs sector may not consist purely of DPs, as assumed in earlier analytical work.
Instead, it is generally composed of both the DP and the dark Higgs. 
The ratio of DP and dark Higgs components is set by post-resonance redistribution processes. 
In the efficient resonance regime considered here, the DP is heavier than the dark Higgs, and an inefficient post-resonance redistribution typically leaves a relic dominated by DPs. Efficient conversion into the lighter dark Higgs can make the two components comparable or, in the limiting case, leave a predominantly scalar relic. 
The late-time dark sector is therefore generically multicomponent, which can have observational consequences: multi-component bosonic dark matter has been shown to affect the core and halo properties of galaxies~\cite{vanDissel:2023vhu}.

Another key result of this work is that the GW signal, relic abundance, and dark-sector mass scale are highly correlated. 
Assuming that the resonantly produced dark sector is responsible for $100\%$ of the observed dark matter abundance, $\Omega_{\rm X}h^2\simeq0.12$, the peak amplitude and the peak frequency of the GW signal map to the initial dark-sector energy fraction $\alpha_{\Phi,*}$, and dark-sector mass scales, respectively.

Over the frequency range $f_{\rm {GW}}\sim {\rm nHz}-100\,{\rm Hz}$ potentially within reach of current and foreseeable GW experiments, this correlation points to ultralight DPs with masses of $10^{-18},{\rm eV}-\mu{\rm eV}$.
The GW probe proposed in our work is complementary to existing non-gravitational searches for light DPs over this mass range, including direct searches such as SuperMAG, SNIPE Hunt, NASDUCK and DMRadio~\cite{Caputo:2026pdw, Friel:2024lti,Silva-Feaver:2016qhh,Caputo:2021eaa,Sulai:2023zqw,Barir:2026awl}.

Several directions are worth exploring in future work. For example, a more detailed treatment of post-trapping number-conserving scatterings would sharpen predictions for the final composition of DPs and dark Higgs. 
It would also be useful to explore additional dark-sector interactions that could deplete the dark Higgs population beyond the minimal case of a quartic coupling. 
Finally, the astrophysical imprints of a dark matter sector composed of ultralight dark Higgs and DP fields merit further investigation. 

This work advances our understanding of a well-motivated nonlinear dark matter production mechanism in the early Universe and opens several avenues for exploring its cosmological and astrophysical implications.

\medskip

\prlsection{Achkowledgements.}YC is supported by the US Department of
Energy under award number DE-SC0008541.
P.~S. is grateful to John T. Giblin, Jr, for insightful discussion on gauge field simulations.
P.~S. is supported by an appointment at the APCTP through the Science and Technology Promotion Fund and Lottery Fund of the Korean Government. 
This was also supported by the Korean
Local Governments-Gyeongsangbuk-do Province and Pohang City.
P.~S. was supported by Grant-in-Aid for Scientific Research (B) under Contract No.~23K25873 (23H01177).
EIS is grateful for support from the Jeff \& Gail Kodosky Endowed Chair at the University
of Texas, Austin, and the U.S. Department of Energy, Office of Science, Office of High Energy Physics program under Award Number DESC-0022021.
Numerical computation in this work was carried out at the Yukawa Institute Computer Facility.

\bigskip

\bibliography{biblio}

\clearpage 

\medskip

\clearpage
\newpage

\onecolumngrid
\setcounter{secnumdepth}{3}
\setcounter{equation}{0}
\setcounter{figure}{0}
\setcounter{table}{0}
\setcounter{page}{1}
\makeatletter
\renewcommand{\theequation}{S\arabic{equation}}
\renewcommand{\thefigure}{S\arabic{figure}}
\renewcommand{\bibnumfmt}[1]{[#1]}
\renewcommand{\citenumfont}[1]{#1}
\pagestyle{plain}

\begin{center}
\Large{\textbf{Seeing in the Dark: Gravitational Waves from Resonant Production of Dark Photon Dark Matter}}\\
\medskip
\textit{Supplemental Material}\\
\medskip
{Yanou Cui, Pankaj Saha and Evangelos I. Sfakianakis }
\end{center}

\section{\label{sec:RDsim}Details of the lattice simulation}
In this section, we describe the details of our lattice simulation, in which we evolve the scale factor self-consistently in an expanding, radiation–dominated (RD) universe for our system consisting of a charged complex scalar field minimally coupled to an Abelian gauge field.
We used a modified version of the lattice simulation code introduced in~\cite{Saha:2026cay}.
We focus mainly on the differences between our implementation and standard lattice codes in the literature~\cite{Deskins:2013dwa,Adshead:2015pva,Adshead:2018doq,Figueroa:2021yhd}.
The background spacetime is spatially flat FRW,
\begin{equation}
  ds^2 = a^2(\tau)\left(-d\tau^2 + d\mathbf{x}^2\right),
\end{equation}
where $\tau$ is conformal time and $a(\tau)$ the scale factor. 
In a purely radiation–dominated universe filled by a homogeneous perfect fluid with equation of state $w = 1/3$, the Universe expands as $a(\tau)\propto\tau$ and $\mathcal{H}=1/\tau$. 
Rather than using this scale-factor ansatz by hand, we model an explicit homogeneous radiation component with energy density $\rho_{\mathrm{R}}(\tau)$ and evolve the scale factor self-consistently from the lattice fields. 
At each time step, we compute the volume-averaged energy density and pressure of the scalar and gauge fields, $\rho_{\mathrm{fields}}$ and $p_{\mathrm{fields}}$, and add a homogeneous
radiation fluid with $p_{\mathrm{R}} = \rho_{\mathrm{R}}/3$. 
The total energy density and pressure,
\begin{equation}
  \rho_{\mathrm{tot}} = \rho_{\mathrm{fields}} + \rho_{\mathrm{R}} \, ,
  \qquad
  p_{\mathrm{tot}} = p_{\mathrm{fields}} + \frac{1}{3}\rho_{\mathrm{R}} \, ,
\end{equation}
are then inserted into the Friedmann and Raychaudhuri equations in conformal time. 
We choose the initial value of $\rho_{\mathrm{R}}$ such that $\rho_{\mathrm{R}} \gg \rho_{\mathrm{fields}}$, so that the background expansion is very close to that of an ideal RD universe, $a(\tau)\propto\tau$, while still allowing the scalar–gauge system and the produced gravitational waves to backreact consistently on the expansion at the level implied by the Friedmann constraint. 
This also guarantees that ratios such as $\rho_{\mathrm{GW}}/\rho_{\mathrm{tot}}$ are computed with a dynamically consistent $\rho_{\mathrm{tot}}$ rather than with a fixed, externally prescribed background.
\par
The matter sector consists of a complex scalar field $\Phi$ with gauge coupling
$g_{_{\!A}}$ for the $U(1)$ gauge field $A_\mu$ (we set the gauge charge $Q_{\!A} = 1$),
\begin{equation}
  \Phi = \frac{1}{\sqrt{2}}\bigl(\phi_1 + i \phi_2\bigr) \, ,
\end{equation}
where $\phi_1$ and $\phi_2$ are real scalar fields. 
We work in the temporal gauge,
\begin{equation}
  A_0(\tau,\mathbf{x}) = 0 \, ,
\end{equation}
so that the dynamical gauge degrees of freedom are the spatial components $A_i(\tau,\mathbf{x})$, and the Gauss constraint plays the role of a
time-independent constraint equation. 
All equations of motion are written in conformal time, so that spatial gradient terms take their Minkowski–space form while the expansion enters through powers of $a(\tau)$ and the
conformal Hubble parameter
\begin{equation}
  \mathcal{H} \equiv \frac{a'}{a}\,.
\end{equation}
We adopt a semi-compact scheme: the gauge field $A_i$ itself is stored as a real-valued non-compact variable on links, but whenever the scalar couples to the gauge field, we reconstruct the Wilson line
\begin{equation}
U_i(x) = e^{-i g_A A_i(x)\,\dx{x}} \, .
\label{eq:WilsonLineVar}
\end{equation}

The scalar field fluctuations, being sub-Hubble, are set in the standard way using the adiabatic vacuum condition~\cite{Felder:2000hq,Figueroa:2020rrl}, 
while the gauge fields are initialized using the Gauss constraint~\cite{Figueroa:2021yhd}:
\begin{equation}
    \partial_i F^{0i} = a^2J^0\,,
\end{equation}
where the current is given by $J^{\mu}\equiv 2g_A\mathrm{Im}[\Phi^\ast (D^\mu\Phi)]$.

\subsection{\label{sec:gwsLatt}Gravitational waves on lattice}
To extract the gravitational-wave background sourced by the nonlinear field dynamics of our scalar-gauge system, we employ the spectral method of Ref.~\cite{Easther:2007vj}, which evolves the transverse-traceless (TT) metric perturbations in Fourier space, sourced by the TT part of the field anisotropic stress.

Working in the synchronous gauge,
\begin{equation}
  ds^2 = dt^2 - a^2(t)\left[\delta_{ij} + h_{ij}\right]dx^i dx^j,
  \qquad
  h^i{}_i = 0,\quad \partial_i h^i{}_j = 0,
  \label{eq:synchronous_metric}
\end{equation}
the perturbed Einstein equations give the evolution of each Fourier mode
$h_{ij}(\tau,\vec k)$,
\begin{equation}
  h_{ij}'' + 2\mathcal{H}\,h_{ij}' + k^2 h_{ij} = 16\pi G\, a^2(\tau)\,\Pi_{ij}^{\rm TT}(\tau,\vec k),
  \label{eq:hij_k}
\end{equation}
where primes denote derivatives with respect to conformal time $\tau$, $\mathcal{H}=a'/a$, and $\Pi_{ij}^{\rm TT}$ is the transverse-traceless part of the anisotropic stress. 
The source is built from the field gradients,
\begin{equation}
  \Pi_{ij} = \partial_i\phi\,\partial_j\phi
    + \left[\text{gauge-field contributions}\right],
  \label{eq:Pi_source}
\end{equation}
summed over all dynamical fields, and is projected onto its TT part in Fourier space,
\begin{equation}
  \Pi_{ij}^{\rm TT}(\vec k) = \Lambda_{ij,lm}(\hat k)\,\Pi_{lm}(\vec k),
  \qquad
  \Lambda_{ij,lm} = P_{il}P_{jm} - \tfrac12 P_{ij}P_{lm},
  \quad
  P_{ij} = \delta_{ij} - \hat k_i\hat k_j.
  \label{eq:TT_projector}
\end{equation}
Projecting the source~---~rather than imposing the TT constraints on $h_{ij}$ during evolution~---~guarantees that the evolved perturbations remain transverse-traceless, since the TT subspace is preserved by the wave operator in Eq.~\eqref{eq:hij_k}. 
In practice we evolve the six independent components of the symmetric tensor $h_{ij}$ with a fourth-order Runge--Kutta integrator; the TT source ensures that only the two physical polarizations are excited, which we verify by monitoring $\hat k_i h_{ij}(\vec k)$ and $h^i{}_i(\vec k)$ throughout the evolution.

The present-day gravitational-wave energy spectrum is
\begin{equation}
  \frac{d\Omega_{\rm{GW}}}{d\ln k}
  = \frac{1}{\rho_{\rm{crit}}}\frac{d\rho_{\rm{GW}}}{d\ln k},
  \qquad
  \frac{d\rho_{\rm{GW}}}{d\ln k}
  = \frac{k^3}{8\pi G\,L^3}\sum_{i,j}\big|h_{ij}'(\tau,\vec k)\big|^2,
  \label{eq:Omega_GW}
\end{equation}
where $L$ is the comoving box size, redshifted from the end of the simulation to today using
\begin{equation}
  \Omega_{\rm{GW}}h^2 = \Omega_{\rm{rad},0} h^2 \left(\frac{g_{*,0}}{g_{*,e}}\right)^{1/3}
    \left.\frac{d\Omega_{\rm{GW}}}{d\ln k}\right|_{a_e},
  \label{eq:Omega_GW_today}
\end{equation}
with $\Omega_{\rm{rad},0}h^2$ the present radiation density and $a_{\rm{e}}$ the scale factor at the end of the simulation.

The corresponding present-day frequency is
\begin{equation}
f_0 = \frac{k}{2\pi a_0} = \frac{k_{\rm {phys,*}}}{2\pi}
\frac{T_0}{T_*}
\left(\frac{g_{s,0}}{g_{s,*}}\right)^{1/3} \, ,~
k_{\rm {phys,*}}\equiv \frac{k}{a_*} \, .
\label{eq:f0_GW_sup}
\end{equation}

\section{\label{sec:relic}Relic abundance and cosmological viability}

In this section, we derive the compact relic abundance formula used in the main text. 
Parametric resonance converts the coherent dark Higgs condensate into dark vectors $A$, radial scalars $h$, and possibly a small residual coherent component, with abundances and spectra fixed by the nonlinear evolution captured by our lattice simulations. 
Since the dark sector is sequestered from the Standard Model (SM) and does not thermalize, we introduce no dark-sector temperature. All yields are instead normalized to the conserved visible entropy density, $s_{\rm{SM}}a^3={\rm{const}}$, which is unchanged because dark production injects no Standard Model entropy.

\medskip 

With $V(\Phi)=\lambda_\Phi(|\Phi|^2-v^2/2)^2$ and $\Phi=(\phi/\sqrt2)e^{i\theta}$, the vacuum masses are
\begin{equation}
m_h=\sqrt{2\lambda_\Phi}\,v \, ,\qquad m_A=g_Av,\;\text{with}\;
q\equiv\frac{g_A^2}{\lambda_\Phi}=\frac{2m_A^2}{m_h^2} \, ,
\label{eq:masses}
\end{equation}
so that the resonance parameter $q$ fixes the mass hierarchy. 
Here $\varphi\equiv \sqrt{2}|\Phi|$ is the canonically normalized radial
amplitude, so that in the large-field quartic regime
\(\rho_{X,*}\simeq \lambda_\Phi\varphi_*^4/4\).
The lattice data are specified at a reference epoch (subscript $\ast$) deep in the quartic regime, $\varphi_\ast\gg v$, where $\rho_{{\rm X},\ast}$ redshifts as radiation. 
The initial dark sector energy fraction and the dark-to-visible energy ratio,
\begin{equation}
\alpha_{\Phi,\ast}\equiv\frac{\rho_{{\rm X},\ast}}
{\rho_{{\rm{SM}},\ast}+\rho_{{\rm X},\ast}}\, ,\qquad
\mathfrak{R}_{\Phi,\ast}\equiv\frac{\rho_{{\rm X},\ast}}{\rho_{{\rm{SM}},\ast}}
=\frac{\alpha_{\Phi,\ast}}{1-\alpha_{\Phi,\ast}} \, \qquad,
\label{eq:alpha}
\end{equation}
are the primary inputs. Expressing $\mathfrak{R}_{\Phi,\ast}$ in terms of $\rho_{X,*}\simeq \lambda_\Phi\varphi_*^4/4$ and $\rho_{{\rm {SM}},\ast}=\tfrac{\pi^2}{30}g_\ast T_\ast^4$, we find the useful relation that relates the lattice inputs to the cosmological variables:
\begin{equation}
\frac{T_\ast}{\sqrt{\lambda_\Phi}\,\phi_\ast}
=\left[\frac{30}{4\pi^2 g_\ast}\right]^{1/4}
\lambda_\Phi^{-1/4}\mathfrak{R}_{\Phi,\ast}^{-1/4} \, .
\label{eq:conversion}
\end{equation}
Every yield below inherits its $\lambda_\Phi^{-1/4}\mathfrak{R}_{\Phi,\ast}^{-1/4}$
structure from Eq.~\eqref{eq:conversion}.

When nonlinear energy transfer reaches saturation, the comoving energy is partitioned into vector, scalar-fluctuation, and residual-condensate fractions $\eta_A,\eta_h,\eta_0$ (with $\sum_{i\in\{A,h,0\}}\eta_i=1$), and the mean energy and momentum per quantum are parameterized by dimensionless  coefficients 
\begin{equation}
\frac{\tilde{\epsilon}_i}{a_\ast}=\kappa_i^{E}(q)\sqrt{\lambda_\Phi}\phi_\ast \, ,
\qquad
\frac{\tilde{p}_i}{a_\ast}=\kappa_i^{p}(q)\sqrt{\lambda_\Phi}\phi_\ast \, ,
\label{eq:kappa}
\end{equation}
where $\kappa_i^{E}$  sets the mean energy per quantum and therefore the
abundance, while $\kappa_i^{p}$ sets the typical momentum and therefore
the coldness of the relic.
For a narrow ultrarelativistic spectrum $\kappa_i^E\simeq \kappa_i^p$. 
 Entropy conservation in the visible sector then gives the yields
\begin{equation}
Y_{i}=\frac{n_i}{s_{\rm{SM}}}
=C_Y\,\frac{\eta_i}{\kappa_i^{E}(q)}\,
\lambda_\Phi^{-1/4}\mathfrak{R}_{\Phi,\ast}^{3/4} \, ,
\qquad i=A,h \, ,
\label{eq:Yi}
\end{equation}
with the model-independent prefactor
\begin{equation}
C_Y=\frac{3}{4}\frac{g_\ast}{g_{\ast s}}
\left[\frac{30}{4\pi^2 g_\ast}\right]^{1/4}\simeq0.22
\qquad(g_\ast=g_{\ast s}=100) \, .
\label{eq:CY}
\end{equation}
If the dark Higgs condensate is not completely fragmented, the surviving zero mode contributes $Y_{h,0}=C_Y\,\eta_0\nu_0\,
\lambda_\Phi^{-1/4}\mathfrak{R}_{\Phi,\ast}^{3/4}$, where the adiabatic invariant
of the quartic oscillator fixes $\nu_0=2\sqrt2\,K/\pi\simeq1.18$ with
$K=\int_0^1\!du/\sqrt{1-u^4}$. Collecting all three possible contributions, the dark-sector yield at the time of Higgs vacuum trapping is
\begin{equation}
Y_N^{\mathrm{vt}}=C_Y\,\Xi_N\,\lambda_\Phi^{-1/4}\mathfrak{R}_{\Phi,\ast}^{3/4} \, ,
\qquad
\Xi_N\equiv\frac{\eta_A}{\kappa_A^{E}}+\frac{\eta_h}{\kappa_h^{E}}
+\eta_0\nu_0 \, ,
\label{eq:YN}
\end{equation}
where a single spectral combination $\Xi_N$ carries all lattice input to the
abundance, and $r_{A}^{{\mathrm{vt}}}=(\eta_A/\kappa_A^{E})/\Xi_N$ is the vector
number fraction. 

\medskip

\emph{Relic abundance after vacuum trapping.}---When the Higgs field settles near the true minimum at
$\varphi=v=\mu/\sqrt{\lambda_\Phi}$, number-changing or
number-redistributing reactions may still reshape the composition of the dark sector.
Their net effect is captured by a yield factor and a final vector fraction,
${\cal D}_N\equiv(Y_A^f+Y_h^f)/Y_N^{\mathrm{vt}}$ and $r_{A}^{f}\equiv
Y_A^f/(Y_A^f+Y_h^f)$, with ${\cal D}_N=1$ for number-conserving dynamics such
as $AA\!\leftrightarrow\!hh$ (open for $q>1/2$). For a stable relic that is
nonrelativistic today,
\begin{equation}
\Omega_{\rm X}h^2=2.742\times10^{8}
\Big[\tfrac{m_A}{\rm {GeV}}Y_A^f+\tfrac{m_h}{\rm {GeV}}Y_h^f\Big] \, ,
\end{equation}
which using Eqs.~\eqref{eq:YN} and $m_h/m_A=\sqrt{2/q}$ becomes the master
relation
\begin{equation}
\Omega_{\rm X}h^2=2.742\times10^{8}\,
{\cal D}_N C_Y \Xi_N\,
\lambda_\Phi^{-1/4}\mathfrak{R}_{\Phi,\ast}^{3/4}
\tfrac{m_A}{\rm{GeV}}\,F_m \, ,
\label{eq:master}
\end{equation}
with the composition factor
\begin{equation}
F_m(q,r_{A}^{f})=r_{A}^{f}+\sqrt{2/q}\,(1-r_A^{f}) \, .
\label{eq:Fm}
\end{equation}
Inverting Eq.~\eqref{eq:master} gives the relic contour,
\begin{equation}
m_A=\frac{\Omega_{\rm X}h^2}
{2.742\times10^{8}\,{\cal D}_N C_Y\Xi_N F_m}\,
\lambda_\Phi^{1/4}
\Big(\tfrac{\alpha_{\Phi,\ast}}{1-\alpha_{\Phi,\ast}}\Big)^{-3/4}
{\rm{GeV}}
\label{eq:contour}
\end{equation}
so that for a given initial dark-sector energy fraction $\alpha_{\Phi,\ast}$ and a fixed $q$ (and hence the same resonance structure), we have $m_A\propto\lambda_\Phi^{1/4}\,{\cal D}_N^{-1}$.
Because the relic mass and the peak frequency [Eq.~\eqref{eq:fpeak_fit}] share the same $\lambda_\Phi^{1/4}$ scaling, a given $\lambda_\Phi$ fixes both simultaneously: the dark photon mass and the gravitational-wave frequency are locked together along the relic contour, and scanning $\lambda_\Phi$ across its viable range moves the  $(m_A,\,f_{\rm peak})$ pair across the full span of gravitational-wave detector frequencies.

\medskip

\emph{Cosmological viability.}---Apart from satisfying the correct relic abundance, $\Omega_{\rm X}h^2\simeq0.12$, a viable point must satisfy several consistency conditions. 
First, while the dark sector remains in the quartic regime at BBN, it contributes to $N_{\mathrm{eff}}$. Evaluated above $e^\pm$ annihilation,
\begin{equation}
\Delta N_{\mathrm{eff}}^{\mathrm{BBN}}\simeq{\frac{43}{7}} \,
\frac{\alpha_{\Phi,\ast}}{1-\alpha_{\Phi,\ast}}\,
\mathcal{G}(T_{\mathrm{BBN}},T_\ast) \, ,
\label{eq:Neff}
\end{equation}
where 
\beq
\mathcal{G}=\frac{g_\ast(T_\ast)}{g_\ast} \left( \frac{g_{\ast s}}{g_{\ast s}(T_\ast)}\right )^{4/3}
\eeq
accounts for visible-sector entropy release between dark matter production and BBN.
Equivalently, neglecting entropy dilution gives $\rho_X/\rho_{\rm{SM}}=(7/43)\Delta N_{\rm{eff}}^{\rm{BBN}}$. 
In the main text, we use this conservative choice, for which $\Delta N_{\rm{eff}}^{\rm{BBN}}\lesssim0.3$ gives $\alpha_{\Phi,*}\lesssim0.05$.

Second, each component with an appreciable final energy fraction must
become non-relativistic before matter-radiation equality.
Writing the radial amplitude at vacuum trapping as
$\varphi_{\rm{vt}}=\chi_{\rm{vt}}v$, the typical post-vacuum trapping velocities
are estimated from the redshifted physical momenta as
\begin{equation}
u_{A,{\rm{vt}}}\sim
\frac{p_{A,{\rm{vt}}}}{m_A}
=
\frac{\chi_{\rm{vt}}\kappa_A^p}{\sqrt q} \, ,
\qquad
u_{h,{\rm{vt}}}\sim
\frac{p_{h,{\rm{vt}}}}{m_h}
=
\frac{\chi_{\rm{vt}}\kappa_h^p}{\sqrt 2} \, .
\end{equation}
After vacuum trapping, the masses are fixed while the physical momenta
continue to redshift as $p_i\propto a^{-1}$. Thus
\begin{equation}
\frac{p_i(T)}{m_i}
=
u_{i,{\rm{vt}}}\,
\frac{T}{T_{\rm{vt}}}
\left[
\frac{g_{*s}(T)}{g_{*s}(T_{\rm{vt}})}
\right]^{1/3}.
\end{equation}
For a component that is still relativistic at vacuum trapping, the temperature at which it becomes non-relativistic is defined by
$p_i(T_{{\rm{NR}},i})/m_i\simeq1$, giving
\begin{equation}
T_{{\rm{NR}},i}
=
\frac{T_{\rm{vt}}}{u_{i,{\rm{vt}}}}
\left[
\frac{g_{*s}(T_{\rm{vt}})}{g_{*s}(T_{{\rm{NR}},i})}
\right]^{1/3}.
\end{equation}
If $u_{i,{\rm{vt}}}\lesssim1$, the component is already
non-relativistic at vacuum trapping. We require
\begin{equation}
T_{{\rm{NR}},i}>T_{\rm{eq}}\simeq0.8\,{\rm{eV}}
\end{equation}
for each component with a non-negligible final energy fraction
$f_i^{\rho,f}$. Points with $T_{\mathrm{eq}}<T_{{\mathrm{NR}},i}
\lesssim1$~keV remain matter-like by equality but may suppress small-scale power; we do not exclude them, because the conventional keV screen~\cite{Dror:2018pdh} is not a sharp bound and the true suppression depends on the full nonthermal phase-space distribution. 
Only the hard $T_{{\mathrm{NR}},i}<T_{\mathrm{eq}}$ region is shaded as an exclusion. The vacuum trapping temperature itself, evaluated along the contour, is $T_{\mathrm{vt}}\propto\chi_{\mathrm{vt}}\Omega_{\rm X}h^2/(\sqrt q\,\Xi_N F_m
{\cal D}_N)\times(1-\alpha_{\Phi,\ast})/\alpha_{\Phi,\ast}$, with the explicit $\lambda_\Phi$ dependence canceling.

Finally, the resonance must have enough time to complete before the oscillation envelope redshifts to the vacuum scale. We use $\varphi_*/v\gtrsim{\cal O}(10\!-\!100)$ as a conservative validity criterion, reflecting the uncertainty in the integrated Floquet growth. This is a condition on the efficiency of the production mechanism rather than on the late-time relic abundance. The region $\phi_\ast/v \lesssim 10$ is marked ``insufficient resonance'' in Fig.~\ref{fig:plotRelic}. 
Along the relic contour, we have $\phi_\ast/v \gg 10$ throughout, so the mechanism operates with a large margin everywhere in the displayed band, and the shaded region never approaches the viable line for DM production.

These conditions are summarized in Fig.~\ref{fig:plotRelic}, which
shows the relic contour $\Omega_Xh^2=0.12$ in the
$(f_{\rm {peak}},m_A)$ plane for the two limiting post-trapping
compositions, together with the shaded regions where the production
mechanism or the resulting relic is not viable. The relic density depends
on how the conserved number is partitioned after vacuum trapping, encoded
in
\begin{equation}
F_m = r_A^{f} + \sqrt{\frac{2}{q}}\left(1-r_A^{f}\right).
\end{equation}
We bound the outcome using two limiting cases: no conversion, where the composition
freezes at its post-simulation value, and maximal conversion, where the
abundance is transferred to the lighter species. The true relic lies
between these contours. 
The dark-radiation bound in Eq.~\eqref{eq:Neff} depends only on $\alpha_{\Phi,*}$. In Fig.~\ref{fig:plotRelic} we therefore show the relic contour for a fixed representative value of $\alpha_{\Phi,*}$, rather than shading a separate dark-radiation-excluded region. The
resonance-duration requirement is shown as the dotted, hatched region.
Finally, the peak frequency $f_{\rm {peak}}\propto\lambda_\Phi^{1/4}$ maps
the quartic coupling onto the gravitational-wave axis as described in the
main text.

\medskip

The abundance calculation described above agrees with  analytic estimates~\cite{Dror:2018pdh}
in the appropriate parametric limits, but the quantities entering the
relic prediction, including $\eta_i$, $\kappa_i^{E,p}$, $\chi_{\rm{vt}}$,
$\mathcal{D}_N$, and $r_A^{N,f}$, are obtained from the nonlinear evolution or kept
as explicit post-vacuum trapping inputs. Residual model dependence from
self-interactions, isocurvature, defects, or portal-induced instability
enters only as consistency conditions on the viable region and does not
alter the abundance relations above.

\end{document}